\documentclass[aps,
twocolumn,
amsmath,amssymb,preprintnumbers]{revtex4}

\usepackage{hyperref}
\usepackage{amsmath} \usepackage{amsfonts} \usepackage{amssymb}
\usepackage{graphicx}

\newcommand{\be}{\begin{equation}}
\newcommand{\ee}{\end{equation}}
\newcommand{\bal}{\begin{align}}
\newcommand{\eal}{\end{align}}
\newcommand{\ba}{\begin{eqnarray}}
\newcommand{\ea}{\end{eqnarray}}

\newcommand{\beq}{\begin{equation}}
\newcommand{\eeq}{\end{equation}}
\newcommand{\beqa}{\begin{eqnarray}}
\newcommand{\eeqa}{\end{eqnarray}}

\newcommand{\bea}{\begin{eqnarray}}
\newcommand{\eea}{\end{eqnarray}}
\newcommand{\beann}{\begin{eqnarray*}}
\newcommand{\eeann}{\end{eqnarray*}}
\usepackage{color}
\usepackage{slashed}

\newcommand*{\mn}{{\mu\nu}}

\def\eq#1{(\ref{#1})}

\newcommand {\apgt} {\ {\raise-.5ex\hbox{$\buildrel>\over\sim$}}\ }
\newcommand {\aplt} {\ {\raise-.5ex\hbox{$\buildrel<\over\sim$}}\ }

\def\s0#1#2{\mbox{\small{$ \frac{#1}{#2} $}}}
\def\0#1#2{\frac{#1}{#2}}

\newcommand{\del}{\partial}

\begin{document}

\title[ ]{Scaling solutions for Dilaton Quantum Gravity}

\author{T. Henz} 
\affiliation{Institut f\"ur Theoretische
  Physik, Universit\"at Heidelberg, Philosophenweg 16, 69120
  Heidelberg, Germany} 

\author{J. M. Pawlowski} 
\affiliation{Institut f\"ur Theoretische
  Physik, Universit\"at Heidelberg, Philosophenweg 16, 69120
  Heidelberg, Germany}

\author{C. Wetterich} 
\affiliation{Institut f\"ur Theoretische
  Physik, Universit\"at Heidelberg, Philosophenweg 16, 69120
  Heidelberg, Germany} 
\begin{abstract}
  Scaling solutions for the effective action in dilaton quantum
  gravity are investigated within the functional renormalization group
  approach. We find numerical solutions that connect ultraviolet and
  infrared fixed points as the ratio between scalar field and
  renormalization scale $k$ is varied. In the Einstein frame the
  quantum effective action corresponding to the scaling solutions
  becomes independent of $k$.  

  The field equations derived from this effective action can be used
  directly for cosmology.  Scale symmetry is spontaneously broken by a
  non-vanishing cosmological value of the scalar field.  For the
  cosmology corresponding to our scaling solutions, inflation arises
  naturally. The effective cosmological constant becomes dynamical
  and vanishes asymptotically as time goes to infinity.
\end{abstract}

\maketitle

\section{Introduction}
There is accumulating evidence that quantum gravity may be
non-perturbatively renormalizable due to the existence of an
ultraviolet fixed point. This scenario of asymptotic safety
\cite{weinberg1979ultraviolet} has been found in four-dimensional
renormalization group investigations \cite{Reuter:1996cp}, based on
functional renormalization for the effective average action
\cite{Wetterich:1992yh,Reuter:1993kw}. Many extensions of the truncation
beyond the simplest Einstein-Hilbert truncation have confirmed the
presence of the ultraviolet fixed point
\cite{Dou:1997fg,Souma:1999at,Reuter:2001ag,Litim:2003vp,Codello:2006in,Machado:2007ea,Codello:2008vh,Fischer:2006fz,Benedetti:2009rx,Eichhorn:2010tb,Manrique:2010am,Donkin:2012ud,Christiansen:2012rx,Rechenberger:2012dt,Dietz:2012ic,Codello:2013fpa,Falls:2013bv,Benedetti:2013jk,Christiansen:2014raa,Christiansen:2015rva,Dietz:2015owa,Demmel:2015oqa,Falls:2015qga,Gies:2015tca,Gies:2016con},
for reviews see
\cite{Niedermaier:2006wt,Percacci:2007sz,Litim:2011cp,%
Reuter:2012id,Nagy:2012ef}.
Similar ideas have been explored in other quantum field theories, see,
e.g., \cite{Gies:2009hq,Braun:2010tt,Litim:2014uca}.  Interactions
with matter have been included
\cite{Dou:1997fg,Folkerts:2011jz,Harst:2011zx,Dona:2013qba,Oda:2015sma,Meibohm:2015twa,Eichhorn:2016esv},
with emphasis on scalar matter in
refs. \cite{Narain:2009fy,Narain:2009gb,Percacci:2015wwa,Dona:2015tnf,Labus:2015ska}. Additional
arguments in favor of asymptotic safety arise from different
approaches to quantum gravity \cite{Hamber:2009mt, Ambjorn:2009ts}.

In the presence of an ultraviolet fixed point the flow of couplings
can be extended to the limit where the renormalization scale $k$ goes
to infinity. Within functional renormalization, observable quantities
are obtained in the opposite limit $k\to 0$. Furthermore, the gravitational
quantum field equations relevant for cosmology arise for
$k\to 0$. It is therefore important to find smooth trajectories from
the ultraviolet to the infrared limit as $k$ decreases to zero
\cite{Donkin:2012ud,Christiansen:2012rx, Nagy:2012rn}.

Gravity coupled to a scalar field offers the interesting perspective
that a realistic scale-symmetric theory of gravity and particle
physics can be formulated if the Planck mass is given by a scalar
field $\chi$ \cite{Fujii:1982ms,Wetterich:1987fm,Shaposhnikov:2008xi}. In
the absence of explicit mass scales the scale $k$ can only be compared
to $\chi$, with IR-limit $k/\chi\to 0$. If cosmological solutions for
$\chi$ approach an infrared fixed point with exact scale symmetry the
\textquotedblleft dilatation anomaly\textquotedblright\ close to the
fixed point can give rise to dynamical dark energy or quintessence
\cite{Wetterich:1987fm}. The crossover between an ultraviolet and
infrared fixed point can connect inflation and present dynamical dark
energy, both described by the same cosmon field
\cite{Wetterich:2014gaa}. Realistic cosmology is found in a picture where
the universe is not expanding during radiation and matter domination
\cite{Wetterich:2013aca} and the big bang singularity is absent
\cite{Wetterich:2014zta}.

In the functional renormalization group approach to quantum gravity
the system of a scalar field coupled to gravity was first studied in
ref. \cite{Narain:2009gb,Narain:2009fy}. The existence of a \textquotedblleft
global scaling solution\textquotedblright\ for all $k$ and $\chi$,
with general scalar potential and scalar-field dependent coefficient
of the curvature scalar, was investigated in
ref. \cite{Henz:2013oxa}. Recent advances towards a global scaling
solution have been made in \cite{Percacci:2015wwa,Borchardt:2015rxa}.

In the present work we derive for the first time candidates for
global scaling solutions in dilaton gravity. These solutions are
obtained from a qualitatively improved approximation to the effective
action in comparison to those used in previous works: Firstly, we include a $\chi$-dependent
coefficient of the scalar kinetic term, called the kinetial.  This
closes a systematic derivative expansion in the second order of
derivatives. The second necessary improvement concerns the
computation of dynamical correlation functions on the basis of the
expansion scheme put forward in refs.
\cite{Christiansen:2012rx,Christiansen:2014raa,Christiansen:2015rva},
for gravity-matter systems see refs.
\cite{Meibohm:2015twa,Meibohm:2016mkp,Eichhorn:2016esv}. 
  This goes beyond the standard background
field approximation. With its relation to the constraints of
diffeomorphism symmetry in a gauge-fixed setting it is at the root of
background independence, for discussions see e.g. ref.
\cite{Litim:2002ce,Folkerts:2011jz,Christiansen:2015rva,Meibohm:2015twa}.
The expansion around a flat background as well as a vertex
construction allow us to compute the running of the kinetial as well as to
disentangle fluctuating and background fields.

\section{Setup}
%\medskip\noindent {\em Setup.}  
We aim at finding global fixed point
solutions for the effective action of dilaton gravity
\begin{align}
\label{eq:action} 
\Gamma =\int_x \sqrt{g} \left( V(\chi^2)-\frac12 F(
  \chi^2)\,R+\frac{1}{2} K( \chi^2) g^{\mu\nu}
  \partial_\mu\chi\partial_\nu\chi \right)\,,  
\end{align}
where $\int_x = \int d^4 x$.
The three functions $V$, $F$ and $K$ depend on a scalar field $\chi$
and the renormalization scale $k$. For a scaling solution, the dimensionless functions
$V/k^4$, $F/k^2$ and $K$ only depend on the dimensionless ratio
$y=\chi^2/k^2$. For fixed $k=\mu$ the effective action \eq{eq:action}
constitutes a model of variable gravity, for which the cosmology is
discussed in detail in \cite{Wetterich:2013jsa}. In this work we extract
the scale dependence of the functions $V$, $F$ and $K$ from the
functional renormalization group. This translates directly to the
$\chi$-dependence of these functions and therefore to the field
equations relevant for cosmology.
We work with dimensionless functions and fields.

To derive the flow equations, we consider 
$\Gamma[\bar g,\bar\chi; g,\chi]$.
Here,
$\bar g$, $\bar\chi$ are background
fields and $h=g-\bar g$, $\delta \chi = \chi - \bar \chi$ are the dynamical fluctuation fields.
While the occurrence of the background
metric is inherent to any gauge-fixed approach to quantum gravity, the
dependence on the dilaton background field is only introduced via the
regulator term, see refs. \cite{Litim:2002hj,Dietz:2015owa}.
The identification $\bar g = g$, $\bar \chi = \chi$ eliminates generalized gauge fixing terms and results in the gauge invariant effective action $\Gamma[g,\chi]$.
We are interested in the scaling solution for $\Gamma[g,\chi]$. The flows of $V$, $F$ and $K$ are extracted from the flow of the two point functions for the fluctuating fields $h$ and $\delta \chi$ \cite{Christiansen:2014raa,Christiansen:2015rva}.
We work in deDonder
gauge and neglect the ghost contributions, as they do
not couple directly to the dilaton field,
as well as some subleading terms in the $k$-dependence of the regulator. 
We perform a systematic expansion in powers of $h$ to disentangle contributions from background and fluctuating fields. Accordingly, we compute the flow
equations for $V$ and $F$ via the momentum independent and dependent
part of the flow of the transverse-traceless graviton $2$-point
function, respectively. Moreover, we use the momentum-dependent part
of the scalar $2$-point function for the flow of $K$, expanding around
flat space \cite{Christiansen:2014raa}. 
The full equations are too long to be displayed here.

\section{Large field limit}

\begin{figure*}[t]
\includegraphics[width=\columnwidth]{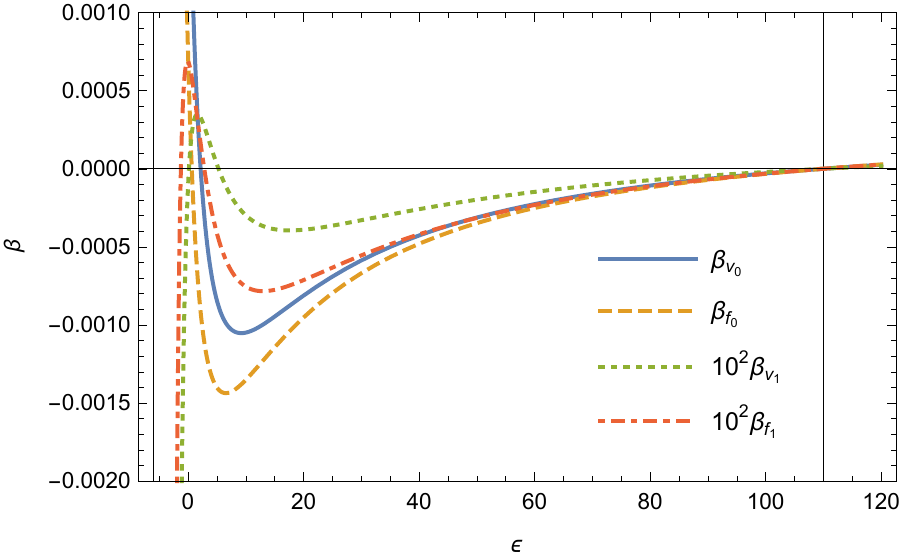}
\includegraphics[width=0.95\columnwidth]{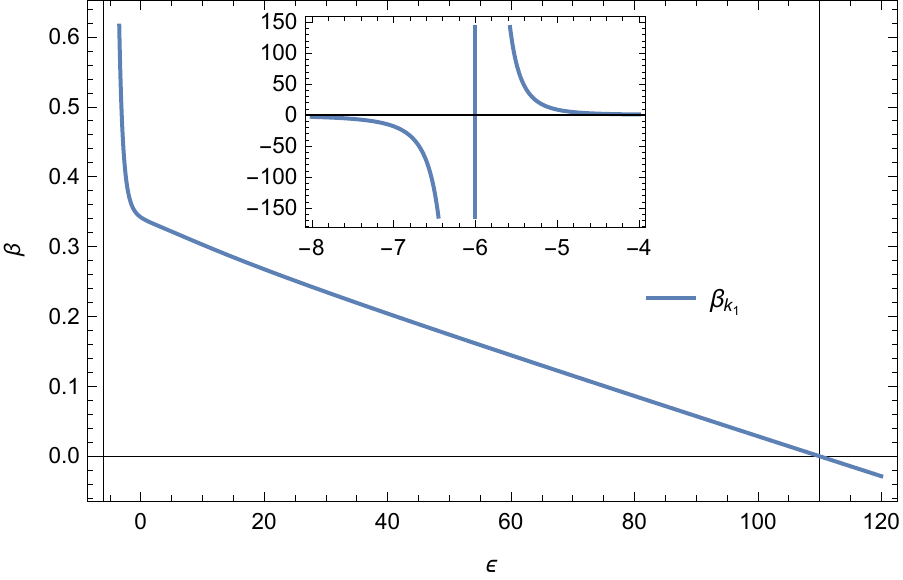}
\caption{Flowing couplings. The $\beta$-functions
  $v_0,f_0,v_1,f_1,k_1$ are shown as functions of $\epsilon$, with other couplings kept fixed at values given by eq. \eqref{eqn:fpsol}. The simultaneous zero at $\epsilon=\epsilon_0=109.97$, as well as the
  pole at $\epsilon=-6$ are clearly visible.}
\label{betafcts}
\end{figure*}
We first consider large $y$, which for finite field $\chi$ is
equivalent to sending the renormalization scale $k\to 0$ and thus
constitutes the infrared limit of dilaton gravity. For fixed $k$ this is the limit $\chi \to \infty$. 
One possible IR-limit implies weak gravity with $F\sim y$. For $\chi\to\infty$ the
effective gravitational constant goes to zero and the gravitational
degrees of freedom decouple. Furthermore, if $K$ and $V/k^4$ approach
constants the asymptotic behavior in the scalar sector is a free
theory. This type of IR fixed point has a very simple physical content: a free massless scalar and a free graviton, with vanishing gravitational coupling. We investigate the scaling solutions connected to this fixed point.

For finite $y$ we expand $V$, $F$ and $K$ in inverse powers of
$y$. For a free scalar field with $F=\xi y$ and $K=K_0$, a rescaling
of $\chi$ multiplies $\xi$ and $K_0$ with the same factor. In
consequence, only the ratio $\epsilon=K_0/\xi$ appears in the flow
equations. More explicitly, for large $y$ we make the ansatz 
 \begin{align}\nonumber 
	 V &= \sum_{j=0}^a \frac{v_j y^{-j}}{j ! \xi^j}\,,\qquad 
 	 F = \xi y + \sum_{j=0}^b \frac{f_j y^{-j}}{j ! \xi^j} ,\, \\[2ex] 
 	 K &= \xi \epsilon +\sum_{j=1}^c \frac{k_j y^{-j}}{j ! \xi^{j-1}}\,.
 \label{largefieldexp}	 \end{align}
Switching to $x=1/(\xi y)$ and evaluating the flow at fixed $x$, one
finds to first order in powers of $x$ the set of flow equations
\begin{align}\nonumber 
  \del_t V & = -4 V -2 x \frac{\del V}{\del x} + \frac{A_V}{B^3} + \frac{C_V}{B^4} x\,, \\[2ex]
  \nonumber
  \del_t F & = -2 F -2 x \frac{\del F}{\del x} + \frac{A_F}{B^3} + \frac{C_F}{B^4} x\,,\\[2ex]
  \del_t K & = -2 x \frac{\del K}{\del x} + \frac{C_K}{B^4} x\,,
\label{eqn:betafcts}\end{align}
with $\del_t = k\del_k$ and 
\begin{align}
\label{eqn:defB}
B=\epsilon+6= \frac{K_0}{\xi}+6.
\end{align}
The terms $-4V$ and $-2F$ reflect the dimensionality of $V$ and $F$,
while the terms $-2x \del V /\del x$, $-2x \del F /\del x$ and $-2x
\del K /\del x$ result from translating the flow to fixed $x$. The
coefficients $A_i$ and $C_i$ are computed from the one loop form of
the flow equation for the effective action. They read
\begin{align} \label{eqn:betafctscoeff} A_V=& \frac{1}{192 \pi ^2}\left(9 \epsilon
    ^3+82 \epsilon ^2+612 \epsilon +2760\right)\,, \\[2ex]\nonumber
  C_V=& \frac{1}{2592 \pi^2}\left(1080 f_0 \epsilon ^3-3240 f_0
    \epsilon ^2-62208 f_0 \epsilon \right.\\[2ex]\nonumber
  &\left. +1080 k_1 \epsilon ^2-3240 k_1 \epsilon -62208 k_1+383 v_0
    \epsilon ^4 \right. \\[2ex]\nonumber & \left.+5004 v_0 \epsilon
    ^3+51120 v_0 \epsilon ^2+278208 v_0 \epsilon +382320
    v_0\right),\\[2ex]\nonumber A_F=& \frac{1}{3456 \pi ^2}\left(-253
    \epsilon ^3-6094 \epsilon ^2-36240 \epsilon -51840\right),
  \\[2ex]\nonumber C_F=& \frac{1}{5184 \pi ^2}\left(2148 f_0 \epsilon
    ^3-2916 f_0 \epsilon ^2-98712 f_0 \epsilon
    +\right. \\[2ex]\nonumber & \left. 2310 k_1 \epsilon ^2-972 k_1
    \epsilon -92880 k_1-345 v_0 \epsilon ^4\right. \\[2ex]\nonumber &
  \left. -23498 v_0 \epsilon ^3-213492 v_0 \epsilon ^2-546552 v_0
    \epsilon \right. \\[2ex]
    &
    \left. -263520 v_0\right)\,,\nonumber \\[2ex]\nonumber
  C_K=& \frac{1}{36 \pi ^2}\left(-\epsilon ^4+90 \epsilon ^3+2079
    \epsilon ^2+12636 \epsilon +26244\right)\,.
\end{align}
The fluctuation contributions do not enter the flow of the leading
terms $F=\xi / x$ and $K=K_0$ such that $\xi$ and $K_0$, and therefore
also $\epsilon$, are arbitrary couplings or \textquotedblleft
integration constants\textquotedblright . The appearance of a free
parameter $\epsilon$ corresponds to the undetermined $\xi$ in an
earlier calculation \cite{Henz:2013oxa} with field independent $K=K_0=1$.

We are interested in fixed point solutions where the left hand side of
\eq{eqn:betafcts} vanishes. The resulting system of
differential equations for the $x$-dependence is closed in every order
in the expansion in $x$. The fixed points for the $x$-independent
terms depend on $\epsilon$ and are given by
\begin{align} 
 v_0=\frac{A_V}{4 B^3}\,,\qquad \ f_0
  =\frac{A_F}{2 B^3}\,.
\end{align} 
Similarly, one has $k_1=C_K/(2 B^4)$. Inserting these values in $C_V$
and $C_F$ yields $v_1$ and $f_1$.

An interesting particular solution arises when one chooses integration constants such that $C_K=0$. This resembles the system investigated in \cite{Henz:2013oxa}.
The condition $C_k=0$ fixes $\epsilon$ to a certain $\epsilon_0$.  For this
value the fixed point solution is given by 
\begin{align}\begin{array}{l@{\quad}l@{\quad}l}
v_0=1.10 \cdot 10^{-3}, & f_0=-3.89\cdot 10^{-3},& \epsilon_0=109.97, \\[2ex] 
v_1=2.32 \cdot 10^{-6}, & f_1=-2.81\cdot 10^{-6},& k_1=0.
\end{array}\label{eqn:fpsol}
\end{align}
It is the only real solution of this type which obeys the stability
condition $\epsilon \geq -6$.

The $\beta$-functions for the couplings $v_0,v_1,f_0,f_1,k_1$
correspond to the coefficients in the expansion of the r.h.s. of
eq. \eqref{eqn:betafcts} in powers of $x$, with
eq. \eqref{largefieldexp} inserted. The $\beta$-functions depend on
the couplings. We plot them in figure \ref{betafcts} for
$v_0,f_0,v_1,f_1$and $k_1$ given by eq. \eqref{eqn:fpsol}, as a
function of $\epsilon$ which is left free. They show a simultaneous
zero at $\epsilon_0=109.97$ as well as the pole at $\epsilon=-6.$ We
point out that our approximation is no longer valid for $B\to 0$.

\section{Global solution}
\begin{figure}[t]
\includegraphics[width=0.9\columnwidth]{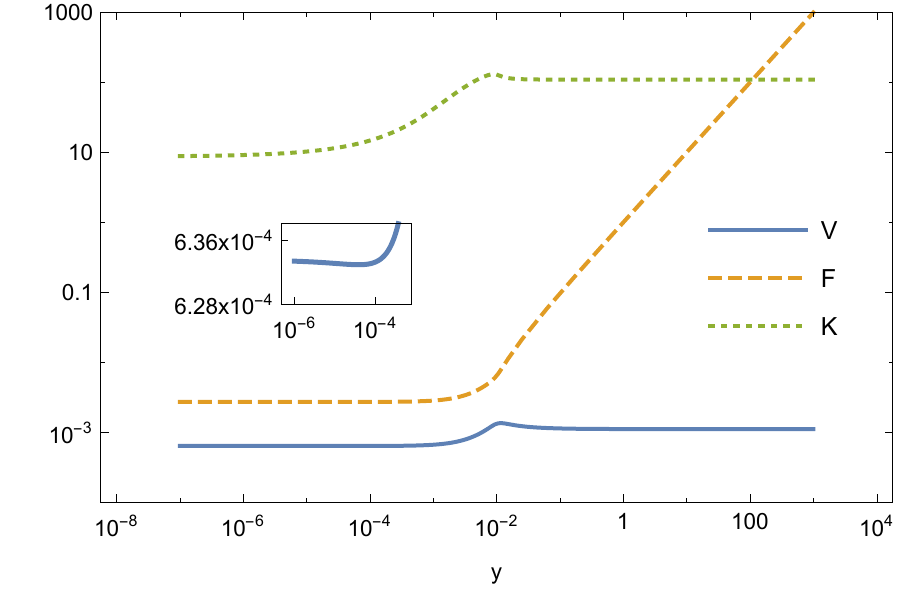}
\caption{Global numerical solution for the functions $V(y)$, $F(y)$
  and $K(y)$ in a double-log plot. The matching with the Taylor
  expansion was carried out at $y=10^5$.}
\label{globalsolall}
\end{figure}

The expansion \eqref{largefieldexp} can be extended to rather high
powers of $x$, and one may perform Pad\'e approximations similar to
ref. \cite{Henz:2013oxa}. It is clear that such expansions will become
unreliable for small $\chi/k$ or large $x$, which we identify with the
ultraviolet limit. In order to connect with the region of large $x$ or small $y$ we need the full flow equations for the dimensionless functions $V/k^4$, $F/k^2$ and $K$. For simplicity, we keep the notation $\{ V,F,K \}$ for the dimensionless quantities. The system of flow equations has been
computed using algebraic algorithms. We have been able to solve numerically the fixed point equations $\del_t V= \del_t F=\del_t K = 0$ for the full functions $V(y),\ F(y)$ and $K(y)$,
except for a range of very small $y$. Initial conditions for the
numerical solution are chosen at large $y$, where the expansion
\eqref{largefieldexp} is valid and can be solved analytically. More
precisely, we have taken the expansion \eqref{largefieldexp} for
$a=b=c=5$ and $y_\text{in}=10^5$. In figure \ref{globalsolall} we
display the numerical solutions for $\epsilon=\epsilon_0$ given by
eq. \eqref{eqn:fpsol}.  Note that there is no point at which $\frac 1
2 F- V \approx 0$, meaning that the previously bothersome singularity
discussed in ref. \cite{Percacci:2015wwa} is not approached by the global
scaling solution.

We have also investigated solutions with values of $\epsilon$ different from $\epsilon_0$. We
were only able to obtain global solutions in the vicinity of
$\epsilon_0$, and only for a finite set of values for
$\epsilon$. While $V$ and $F$ are largely independent of $\epsilon$,
$K$ is rescaled by $\epsilon$ for large $y$, while $\epsilon$ becomes
less and less important for $y\to 0$.

For an estimate of the numerical error we compute the values of the
$\beta$-functions for our numerical solution, which should be
zero. They are normalized to the internal accuracy of the implicit
numerical differential equation solver employed, which was set to $8$
decimal digits. As long as this relative error is smaller than $1$, we can
assume the error to be negligible. This is the case for most parts of
the interval under consideration, with only small, local deviations
due to interpolation errors between the grid points of the numerical
solution. These are inevitable when taking derivatives to compute the
$\beta$-functions from the numerical solutions.

\section{small field limit}
 For an investigation of
the UV-behavior for $\chi/k\to 0$ we first consider the truncation
where $V$, $F$ and $K$ are $\chi$-independent. A scaling solution with
these properties closely resembles pure gravity in the
Einstein-Hilbert truncation \cite{Reuter:1996cp}, except for the presence of
a massless scalar field. In this approximation the coupling between
the scalar and the metric arises purely from the kinetic term. We
include the possibility of a $k$-dependent wave function
renormalization $Z$, or anomalous dimension $\eta=-\del \log Z / \del
\log k$. The wave function renormalization is defined by
$Z=K(y=0)$. At fixed $ y_R = Z y$ only the anomalous dimension $\eta$
(and not $Z$) appears in the flow equations. Here, $\eta$ is
determined by requiring that the $\eta$-dependent $\beta$-functions
maintain for $K( y_R=0)$ an arbitrary $k$-independent value. The
solution
\begin{align} \label{EHsolution} V=0.0006460\,,\quad  F=0.002757\,,
\quad 
\eta=1.6669\,, 
\end{align} 
can be interpreted as the Einstein-Hilbert approximation to our
system. It provides further evidence for the validity of the
Asymptotic Safety Scenario in a truncation extended by a running
kinetial.

In figure \ref{globalsolall} we have only plotted the range $y>10^{-7}$ for our numerical solution. For very small $y$ the
numerics become unstable. This is due to non-analytic behavior which
leads to diverging derivatives and therefore numerical
instabilities. Investigating the first derivative of the numerical
solutions, we find that they diverge as $y^{-\alpha}$ with
$\alpha=\frac 1 2$ for $y\to 0$.  This explains why previous attempts
\cite{Narain:2009fy}, \cite{Percacci:2015wwa}, \cite{Henz:2013oxa} to expand in
integer powers of $y$ were not well suited. A best fit for the three
functions in the limit $y\to 0$, based on data generated from the
numerical solutions for $10^{-7}\leq y \leq 10^{-5}$, is given by
\begin{align}\nonumber 
  V(y)  =& 6.3350\cdot 10^{-4} - 1.7137\cdot 10^{-4}\, y ^{1/2} \\[2ex]\nonumber 
  &+ 8.3172\cdot 10^{-3}\, y + 0.5626\, y^{3/2}\,, \\[2ex]\nonumber 
  F(y)  =& 2.7077\cdot 10^{-3} - 3.7688\cdot 10^{-4} \, y ^{1/2} \\[2ex]\nonumber 
  & + 3.6550\cdot 10^{-3}\, y + 2.4944\, y^{3/2}\,, \\[2ex]\nonumber 
  K(y)  =& 8.7018 + 455.269\, y ^{1/2} \\[2ex]
  & +12940.3\, y + 223285\, y^{3/2}\,.
\label{eqn:bestfits}\end{align}

We observe that the constants approached by $V$ and $F$ in the limit
$y\to 0$ are very close to the Einstein-Hilbert truncation
\eqref{EHsolution}. This is no coincidence: For small $y$ our system
approaches the UV fixed point of asymptotically safe quantum
gravity. The slight numerical differences for $V(0)$ and $F(0)$ may
arise from the small $y$-dependence of $V$, $F$ and $K$ in equation
\eqref{eqn:bestfits}, which has been neglected in the truncation
leading to the values \eqref{EHsolution}. This may have a
stronger influence on the correct behavior of $K(y\to 0)$ and on the
value of the anomalous dimension. Extrapolating
eq. \eqref{eqn:bestfits} to $y\to 0$ would imply $\eta =0$. On the
other hand, it is not excluded that the correct matching of the
proposed scaling solution with the behavior for $y\to 0$ leads to
constraints on the allowed values of $\epsilon$. We are mainly
interested here in the generic IR-behavior which does not depend on
the precise details of the extreme limit $y\to 0$.

\section{Conformal invariants}
\begin{figure}[t]
\includegraphics[width=0.9\columnwidth]{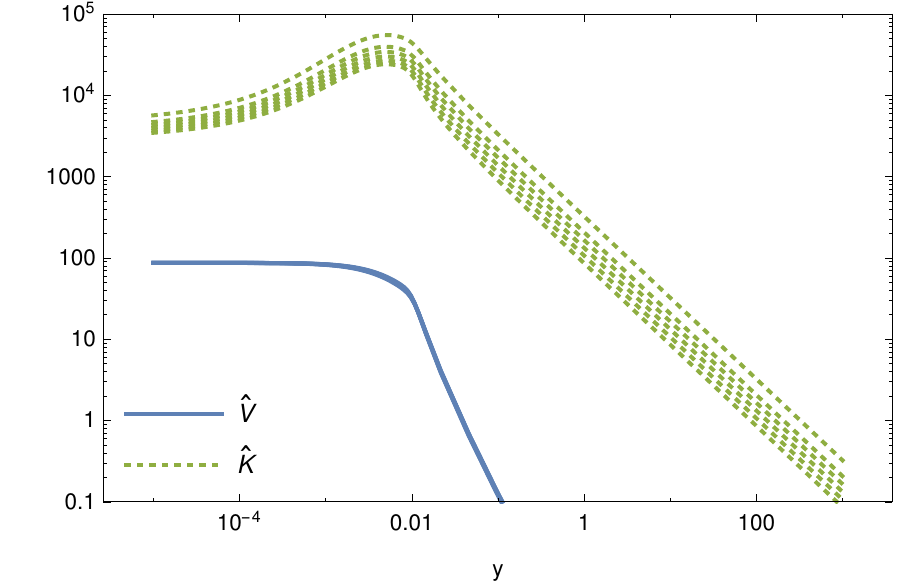}
\caption{Invariants $\hat V$ and $\hat K$ for the fixed point solution
  for different values of $\epsilon$ in the vicinity of $\epsilon_0$.}
\label{VhatKhat}
\end{figure}

Conformal or Weyl transformations of the metric, $g_\mn =
\omega^2(\chi)\tilde g_\mn$, map the set of functions $\{V,F,K\}$ to a
new set $\{\tilde V,\tilde F,\tilde K\}$. By virtue of field
relativity the physical content of a model is specified by the two
invariants under this rescaling, namely \cite{Wetterich:2015ccd}
\begin{align}\hat V = \frac{V}{F^2}, \qquad \hat K = \frac
K F + \frac{6y}{F^2} \left(\frac{\del F}{\del y}\right)^2\,.  
\end{align} 
In figure \ref{VhatKhat} we show these invariants for the numerical
solutions with different values of $\epsilon$. While $\hat V$
shows very little $\epsilon$ dependency, the scaling $\hat K\sim
\epsilon$ is realized only for large $y$. The solutions with different
$\epsilon$ seem not to be equivalent.
The invariants obey 
\begin{align} \lim_{y\to\infty} \hat V= 0\,,\qquad\qquad  \lim_{y\to\infty} \hat K= 0. 
\end{align}
As both $V$ and $K$ go to a constant for $y\to \infty$, while $F$
grows with $\xi y$, we can immediately understand that the ratios
$V/F^2$ and $K/F$ vanish in this limit. Furthermore, we have $y (\del
F / \del y)^2 / F^2=1/y$.  The potential $\hat V$ exhibits a maximum
located at 
\begin{align} \label{eqn:maxVhat} y_\text{max}=0.02\,,\qquad \qquad
  \hat V(y_\text{max})=87\,.
\end{align} 
The vanishing of $\hat V$ for $y\to \infty$ implies that the effective
cosmological constant goes to zero asymptotically for cosmological
solutions where $\chi(t\to\infty)\to\infty$ \cite{Wetterich:1987fm}.

\section{Effective action in the Einstein frame}   
\begin{figure}[t]
\centering
\includegraphics[width=0.9\columnwidth]{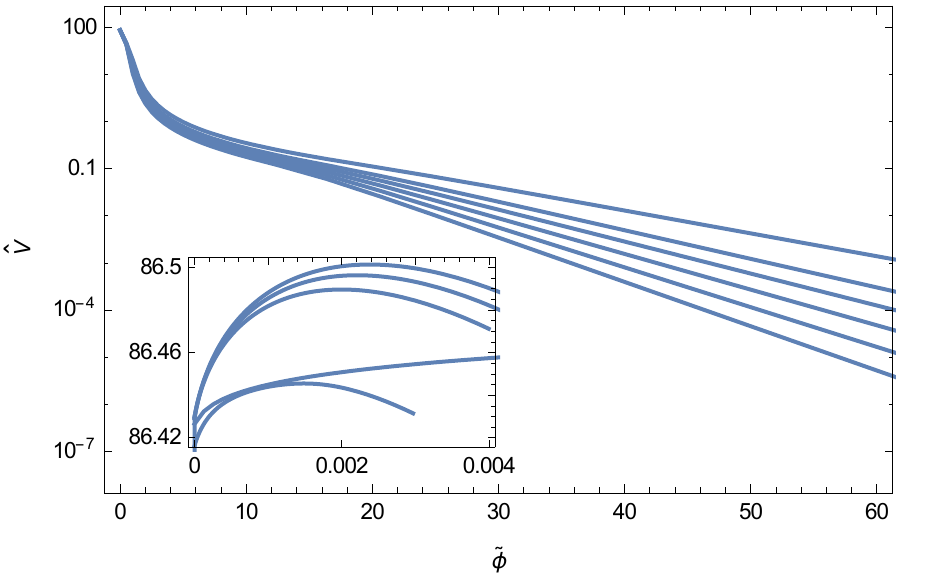}
\caption{Scalar potential $V / M^4$ in the Einstein frame for a standard normalization of
  the scalar field. We show $\hat V$ as a function of $\tilde \phi = \phi/M$ for
  different values of $\epsilon$ around $\epsilon_0$.  Note the
  maximum for small $\tilde \phi$ as well as the exponential tail for
  large $\tilde \phi$.  }
\label{fig:Vhatnormeps}
\end{figure}

Physical features of our system are most readily visible in the
Einstein frame, which is reached by a Weyl scaling leading to $\tilde
F=M^2/k^2$. Further using a rescaling of the scalar field to bring the
kinetic term to standard form yields in the
Einstein frame the effective action 
\begin{align} \label{eqn:EAEinsteinframe} \Gamma =\int_x \sqrt{g}
  \left( M^4 \hat V(\phi)-\frac12 M^2\,R+\frac{1}{2} g^{\mu\nu}
  \partial_\mu\phi\partial_\nu\phi \right)\,.
\end{align}
From the kinetic term
\begin{align}
\label{XA}
\frac{M^2 \hat K}{2 k^2}\del^\mu\chi\del_\mu \chi = \frac 1 2 \del^\mu
\phi \del_\mu \phi\,,\quad \quad \hat K =\frac{16}{\alpha^2 y}\,,
\end{align} 
one infers for constant $\alpha$ \be\label{XB} \frac{1}{
  y^2}=\exp\left(-\frac{\alpha \phi}{M}\right),\quad V=M^4\hat V =
\frac{v_0 M^4}{\xi^2}\exp\left(-\frac{\alpha \phi}{M}\right), \ee with
modifications if $\alpha$ depends on $ y$. We observe that $ y$ is a
function of $\phi/M$, not involving $k$. All memory of $k$ has
disappeared in the Einstein frame. In figure \ref{fig:Vhatnormeps} we
plot the dimensionless potential in the Einstein frame $V/M^4 = \hat
V$ as a function of $\tilde \phi = \phi /M$. For the values of
$\epsilon$ for which numerical solutions could be established the
potential has a maximum for small values of $\tilde \phi$, as shown in
the inset.

\section{Discussion} 
In this note we present for the first time candidates for a global scaling solution for dilaton quantum gravity.
We find several striking
features:\\[-2ex]

(i) For the scaling solution the effective action in the Einstein
frame shows no dependence on $k$. A realistic form of gravity results
directly for the scaling solution, without the need to deviate from
the fixed point. Variation of the effective action
\eqref{eqn:EAEinsteinframe} yields the quantum field equations and one
may discuss cosmological solutions. For fixed $\chi$ the limit $k\to
0$ corresponds to $\phi\to\infty$, but for arbitrarily small $k$ we
can always find values of $\chi$ such that $\phi$ is finite. We may
also consider the possibility that a separate physical cutoff sale
$\mu$ effectively stops the flow for $k\ll\mu$, leading to a deviation
from the scaling solution. A first approximation to the limit $k\to 0$
for this situation effectively replaces $k$ by $\mu$. There could be
additional terms realizing the physical cutoff, such as a mass term in the
potential \eq{eq:action}, $\Delta V = \mu^2\chi^2$.\\[-2ex]

(ii) The scaling solution is invariant under a simultaneous scaling of
fields and $k$. For any non-zero $k$ the effective action
\eq{eq:action} is not scale invariant if only the fields are
rescaled. Standard dilatation or scale symmetry can be recovered for
$k\to 0$. This is realized for our solution where $K$ and $F/y$ go to
constants for $y\to \infty$, while $V/y^2$ goes to zero. Scale
symmetry is spontaneously broken for cosmological solutions with
non-zero $\chi$. Therefore, one will find a massless Goldstone boson, the dilaton, in the limit $y\to\infty$,
$\phi\to\infty$. For finite
large $y$ this corresponds to the cosmon \cite{Wetterich:1987fm} with a
tiny field-dependent mass. Coupling matter fields in a scale invariant
way, e.g. with masses $\sim k\chi$, will lead to a massive particle
spectrum, with constant ratio particle mass / Planck mass.\\[-2ex]

(iii) For the scaling solutions that we have found the potential
asymptotically goes to a constant, $V(y\to\infty)=v_0$. With $F\sim y$
this results in $\hat V\sim y^{-2}$. The potential in the Einstein
frame decreases exponentially for $\phi\to\infty$, according to
equation \eqref{XB}. For cosmological solutions where $\phi\to\infty$
for $t\to\infty$, as realized if one starts on the right side of the
maximum in figure \ref{fig:Vhatnormeps}, the effective cosmological
constant goes to zero asymptotically. This is precisely the mechanism
proposed in the first paper on quintessence \cite{Wetterich:1987fm}. If
nature is characterized by a scaling solution for which $\hat V$
vanishes for $y\to\infty$, the cosmological constant problem is
solved, at least for asymptotic time. Our numerical solution for
$\epsilon=\epsilon_0$ yields $\alpha\approx0.4$. This is too small for
a realistic cosmological scaling solution for late cosmology, which
requires $\alpha \gtrsim 10$ \cite{Wetterich:2014gaa}.\\[-2ex]

(iv) The maximum of the potential for small $\tilde y$ shown in figure
\ref{fig:Vhatnormeps} offers, in principle, the possibility of an
inflationary stage. The end of inflation typically occurs
\cite{Wetterich:2014gaa, Wetterich:2013jsa} once $\hat K y$ drops below a
constant of order one.  A glance at figures \ref{VhatKhat} and
\ref{fig:Vhatnormeps} yields typical values $\hat K y \gtrsim 10$ for
$\tilde y > \tilde y_\text{max}$ (the position of the maximum), with
$\hat K y \approx 10^2$ for $y\gtrsim 10^{-2}$. Inflation presumably
does not end for cosmologies resulting from the scaling solutions.\\[-2ex]

We conclude that the cosmology for the solutions found so far is not
yet realistic. Nevertheless, we emphasize that for a first time we can
directly connect cosmology to scaling solutions in quantum gravity,
without invoking any ad hoc association of $k$ with geometric
quantities. The reason is that no deviation from the scaling solution
is necessary in dilaton quantum gravity. The effective action in the
Einstein frame is independent of $k$, such that the limit $k\to 0$,
which is difficult in other settings, does not need to be performed
explicitly.

It is well possible that other scaling solutions exist beyond those found in this work. The fact that we have not found numerical
solutions for small $\epsilon$ may be related to numerical
instabilities rather than generic absence of such scaling
solutions. In the limit $B\to 0$ the limiting behavior for $y\to
\infty$ is expected to differ from the one discussed here -- the
limits $y\to\infty$ and $B\to 0$ do not commute. A possible scaling
behavior for $y\to \infty$ could be power-like, $V\sim y^{\delta_V},\
F\sim y^{\delta_F},\ K\sim y^{\delta_K}$. This results in $K +
6\delta_F^2 F/y = \hat K F \sim y^{\hat\delta_K}$, where $\hat
\delta_K=\max\left(\delta_K,\delta_F-1\right)$, a natural situation
arising for $\delta_K=\delta_F-1$. The asymptotic vanishing of the
cosmological constant occurs whenever $\delta_V < 2 \delta_F$. Such a
scaling behavior will be much closer to the properties of the infrared
fixed point discussed in ref. \cite{Wetterich:2014gaa}.

The type of scaling solution discussed here is special since we have
required that all three functions $V$, $F$ and $K$ depend only on
$y$. Scaling can also be realized if $y$ is replaced by a renormalized
variable $y_R=Z y$, see the discussion of the small field
limit. Furthermore, in view of possible field transformations we may
require only the invariants $\hat V$ and $y\hat K$ to be functions of
$y_R$. A possible explicit $k$-dependence of the individual functions
$V$, $F$, $K$ then merely reflects the $k$-dependence in the choice of
generalized renormalized fields. By an appropriate choice of a
renormalized metric we may always achieve standard forms as $F=1$ or
$F=y$. Furthermore, by a non-linear dependence of a renormalized
scalar field $\chi_R$ or $\chi$ we can realize standard forms either
of $K$ or of $V$ \cite{Wetterich:2013jsa}. Scaling solutions would then
merely require that the only remaining free function depends only on
the dimensionless ration $\chi_R / k$. Obviously, this condition is
much weaker than the simultaneous scaling of three functions $V$, $F$
and $K$. It will be an interesting question to find out if already a
simple truncation of dilaton quantum gravity admits scaling solutions
that lead to an acceptable cosmology.

\medskip\noindent {\em Acknowledgments.}  We thank Andreas Rodigast
for collaboration on early stages of this project, as well as Aaron
Held for discussions. The authors acknowledge funding by the European
Research Council Advanced Grant (AdG), PE2, ERC-2011-ADG as well as
the German Academic Scholarship Foundation (Studienstiftung des
deutschen Volkes).  \vfill

\bibliographystyle{bibstyle}
\bibliography{FlatDilaton}

\end{document}